\documentclass[10pt,aps,pra,groupedaddress,showpacs,twocolumn,showkeys]{revtex4-1}

\usepackage{amsmath,amssymb}
\usepackage{amsfonts}
\usepackage{bm}
\usepackage{graphicx}
\usepackage{epstopdf}
\usepackage{float}

\begin{document}
\title{Polarizability Expressions for Predicting Resonances in Plasmonic and Mie Scatterers}

\author{R\'{e}mi Colom$^{1,2,*}$, Alexis Devilez$^{1}$, Stefan Enoch$^1$, Brian Stout$^1$, Nicolas Bonod$^{1,*}$}
\affiliation{$^1$Aix Marseille Univ, CNRS, Centrale Marseille, Institut Fresnel, Marseille, France}
\affiliation{$^2$Centre for Ultrahigh-bandwidth Devices for Optical Systems (CUDOS), University of Sydney NSW 2006, Australia}
\email{remi.colom@fresnel.fr}
\email{nicolas.bonod@fresnel.fr}

\begin{abstract}
Polarizability expressions are commonly used in optics and photonics to model the light scattering by small particles. Models based on Taylor series of the scattering coefficients of the particles fail to predict the morphologic resonances hosted by dielectric particles. Here, we propose to use the factorization of the special functions appearing in the expression of the Mie scattering coefficients to derive point-like models. These models can be applied to reproduce both Mie resonances of dielectric particles and plasmonic resonances of metallic particles. They provide simple but robust tools to predict accurately the electric and magnetic Mie resonances in dielectric particles.
\end{abstract}

\maketitle
\section{Introduction and motivations}
Light scattering by subwavelength sized scatterers is a fundamental problem in optics \cite{Hulst1957,Jackson1999,Bohren1983,Mishchenko2002,Novotny2006,tribelsky2006}. The full electromagnetic problem can be solved with the well-known Mie theory that permits to accurately determine the optical response of spherical objects regardless of their size and composition \cite{Mie1908,Gouesbet11}. However, the complexity of the multipolar formalism has motivated the derivation of point-like models providing more insight into the physical processes involved in light scattering. 

Such models have been widely used for example in the case of small metallic particles behaving like electric dipoles and hosting localized surface plasmon resonances (LSPR) \cite{Kelly2003,Novotny2006}. 
In this case, the electric dipolar polarizability $\alpha_{e}$ relating the dipolar moment $\textbf{p}$ to the excitation field $\textbf{E}_{exc}$ is given by $\textbf{p}=\epsilon_{0}\varepsilon_{b}\alpha_{e}\textbf{E}_{exc}$. $\alpha_{e}$ may easily be linked to the dipolar Mie coefficient $a_{1}$ through the relation $\alpha_{e}=i\frac{6\pi}{k^{3}}a_{1}$ \cite{Colom2016,Stout2011}.

Accurate approximations of $\alpha_{e}$ calculated in the long wavelength limit have greatly contributed to extend the understanding of the resonant process responsible for the interesting features of small plasmonic scatterers \cite{Wokaun1982,Meier1983,Kuwata2003,Moroz2009,Francs2009,Albaladejo2010,LeRu2013}. 
Simplified models for metallic particles including the radiative and finite-size corrections were proposed by Moroz \cite{Moroz2009} and by Meier and Wokaun \cite{Meier1983}.
The first is obtained by calculating the power series expansion of the Mie coefficient $a_{1}$ to the third order while the latter is obtained by taking into account the depolarization field. 

High refractive index dielectric subwavelength-sized particles can also resonantly interact with light~\cite{Evlyukhin11,GarciaEtxarri11,Zambrana2015,Savelev2015,decker2016,kuznetsov2016} $via$ the excitation of low order electric and magnetic Mie resonances. Point-like models should then be able to predict electric and magnetic resonances. However, the classical models widely used in plasmonics fail to predict the dipolar electric resonant response of these dielectric scatterers. 
\begin{figure} 
\centering
\includegraphics[width=0.4\textwidth]{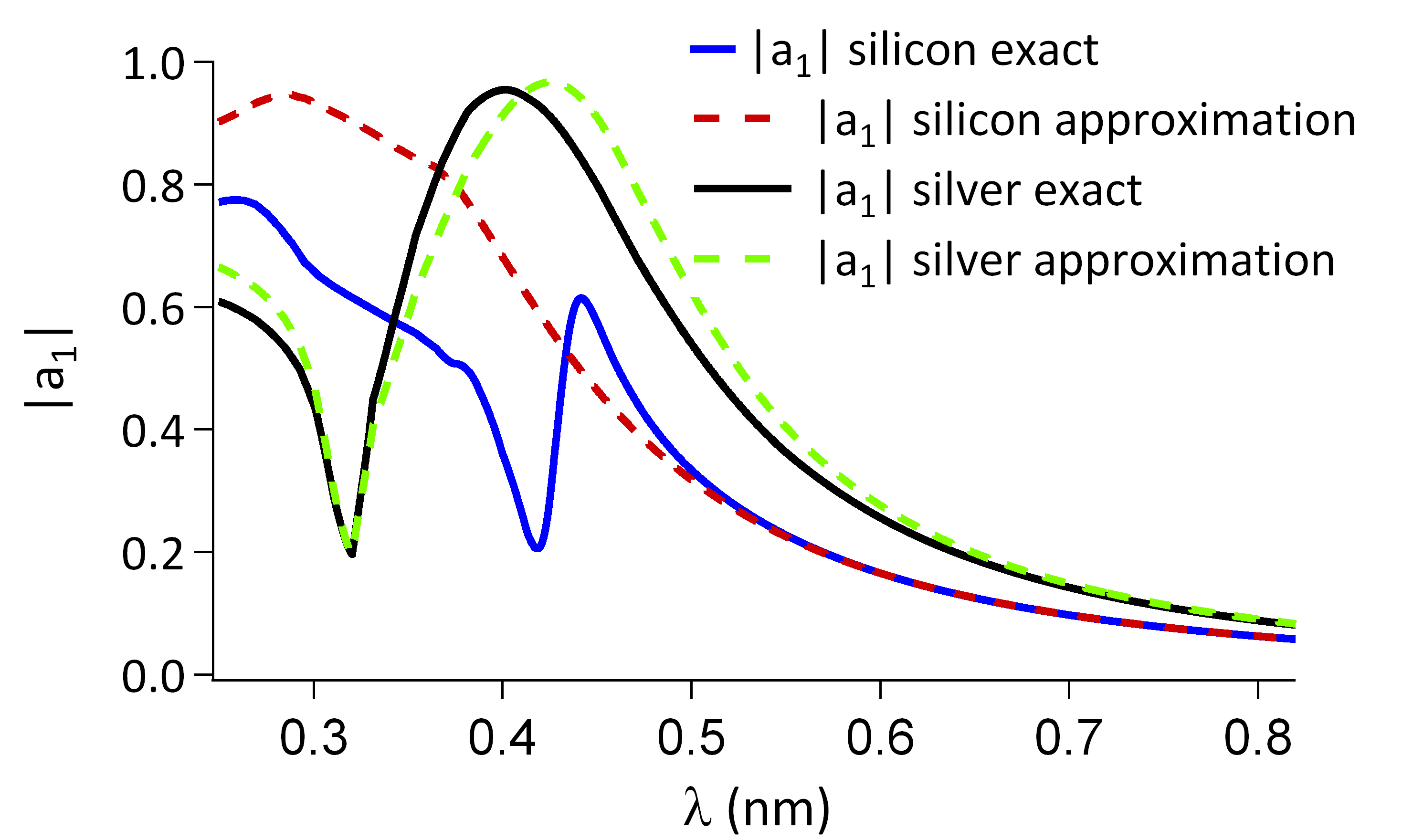}
\caption{(Color Online) Mie coefficient $|a_{1}|$ plotted with respect to the wavelength in the case of a sphere of radius 60 nm made of silicon \cite{Aspnes1983} and silver \cite{Johnson}. Full lines: calculated with the exact expression derived in Eq.1 with $n=1$ with silicon (blue) and silver (black). Dashed lines: approximation $a_{1}^{(T1)}$ in Eq. (\ref{approximations_a1_A3_A4}) taken from \cite{Moroz2009} in red for silicon and green for silver.}
\label{Re_a1_Si_exact_Moroz}
\end{figure}

We illustrate this problem by plotting in Fig.~\ref{Re_a1_Si_exact_Moroz} the real part of the first electric Mie coefficient calculated with the complete Mie theory (full line) and with a Taylor expansion (dashed line) derived up to the 3rd order, Eq. (11) in \cite{Moroz2009}, in the case of a sphere, 120 nm in diameter, made of silver or made of silicon. It is clearly seen that while this expansion does predict the localized surface plasmon resonance around 410 nm, it fails to predict the morphological resonance at 450 nm. 
This issue motivates the development of a generalized point-like model working for both positive and negative dielectrics. Throughout this article, it is demonstrated that the Weierstrass factorization of Bessel functions permits to derive accurate approximations of the Mie coefficients and consequently particle polarizabilities for both dielectric and metallic particles. 
The optical response of spherical scatterers can be accurately described thanks to a set of electric and magnetic Mie coefficients $a_{n}$ and $b_{n}$ \cite{Hulst1957,Bohren1983,Mishchenko2002,Grigoriev2013} which can be expressed: 
\begin{equation}
a_{n}(\varepsilon_{s},z_{0})=\frac{j_{n}(z_{0})}{h_{n}^{(+)}(z_{0})}\frac{\varepsilon_{s} \varphi_{n}^{(1)}(z_{0})-\varphi _{n}^{(1)}(z_{s})}{\varepsilon_{s} \varphi
_{n}^{(+)}(z_{0})-\varphi _{n}^{(1)}(z_{s}),}
\label{Mie_a}
\end{equation}
where $\varepsilon _{s}$ is the relative permittivity : $\varepsilon_{s}=\frac{\epsilon _{s}}{\epsilon _{b}}$ ($\epsilon _{s}$ and $\epsilon_{b}$ being the dielectric permittivities of the sphere and background medium respectively) and $n$ describes the order of the mode. $z_{0}=kR$ is the size parameter of the scatterer, $k$ being the wavenumber $\frac{2\pi}{\lambda}$ and $R$ being the radius of the scatterer considered and $z_{s}=\sqrt{\varepsilon_{s}}z_{0}$ for a non-magnetic material.
The functions $h_{n}^{(+)}$ and $j_{n}$ are respectively the spherical outgoing Hankel functions and the Bessel functions. Finally, the functions $\varphi _{n}^{(1)}$ and $\varphi _{n}^{(+)}$ are reduced logarithmic derivative Ricatti Hankel and Ricatti Bessel functions respectively: 
\begin{equation}\label{express_phi_n}
\varphi _{n}^{(+)}(z) =\frac{[zh_{n}^{(+)}(z)]^{\prime }}{h_{n}^{(+)}(z)}, \varphi _{n}^{(1)}(z) =\frac{[zj_{n}(z)]^{\prime }}{j_{n}(z).}
\end{equation}

Equation (\ref{Mie_a}) remains completely equivalent to the formulation commonly employed in the literature \cite{Bohren1983} and proves to be well adapted for our study. An advantage of this expression is that the magnetic Mie coefficients $b_{n}$ can be obtained from the expression in Eq.~\ref{Mie_a} by replacing the permittivity contrast $\varepsilon_{s}$ by the permeability contrast $\mu_{s}$ (equal to 1 in the case of non magnetic media).

\section{Resonance conditions in subwavelength spheres}\label{section_res_cond}

As a first step toward more general expressions, we propose to determine the resonance conditions (i) graphically and (ii) in the asymptotic limit $z_0 \rightarrow 0$ for any arbitrary made material homogeneous particles.
This methodology will allow us to choose the proper approximations of the special functions appearing in Mie theory in order to derive point-like models valid for metallic and dielectric particles.
We first introduce the K-matrix formulation that will allow us to establish the resonance condition with respect to the $\varphi _{n}^{(1)}(z_s)$ function.
The reactance K-matrix describes the light-scattering by a particle \cite{LeRu2013}. By means of the K-matrix coefficients, one can reformulate the Mie coefficients \cite{Colom2016}:

\begin{eqnarray}\label{an_bn_Kn}
(a_{n})^{-1}&=-i(K^{(e)}_{n})^{-1}+1 \label{an_Kn_e},\\
(b_{n})^{-1}&=-i(K^{(h)}_{n})^{-1}+1 \label{an_Kn_h}.
\end{eqnarray}

where the K-matrix coefficients of a sphere are \cite{Colom2016}: 

\begin{eqnarray}\label{K_n}
K^{(e)}_{n}&=-\frac{j_{n}(z_{0})}{y_{n}(z_{0})}\frac{\varepsilon_{s} \varphi
_{n}^{(1)}(z_{0})-\varphi _{n}^{(1)}(z_{s})}{\varepsilon_{s} \varphi
_{n}^{(2)}(z_{0})-\varphi _{n}^{(1)}(z_{s})}\\
K^{(h)}_{n}&=-\frac{j_{n}(z_{0})}{y_{n}(z_{0})}\frac{\varphi_{n}^{(1)}(z_{0})-\varphi _{n}^{(1)}(z_{s})}{\varphi_{n}^{(2)}(z_{0})-\varphi _{n}^{(1)}(z_{s})}
\end{eqnarray}

where $y_{n}$ are the spherical Neumann functions and $\varphi_{n}^{(2)}(z) =\frac{[zy_{n}(z)]^{\prime }}{y_{n}(z)}$. As the K-matrix is hermitian for non-absorptive particles \cite{LeRu2013,Colom2016}, one can notice that the coefficients $K^{(e)}_{n}$ and $K^{(h)}_{n}$ of a lossless spherical scatterer are real. 
The expression (\ref{an_bn_Kn}) can in fact be seen as a generalization of the energy-conserving formulation of the polarizability $\alpha_{e}^{-1}=\alpha_{n.r.}^{-1}-i\frac{k^{3}}{6\pi}$ \cite{Albaladejo2010, Colom2016}, where $\alpha_{n.r.}=-6\pi K_{1}^{(e)}/k^{3}$ is the non-radiative polarizability, real for lossless scatterers, while the term $-i\frac{k^{3}}{6\pi}$ corresponds to the radiative corrections and is analogue to the $+1$ term in (\ref{an_bn_Kn}).\\ 

\subsection{Definitions of resonances:}
In this study, the light-scatterer interaction will be considered at resonance when one of the Mie coefficients reaches the unitary limit, $i.e.$ when $a_{n}=1$ or $b_{n}=1$ \cite{Colom2016,Videen1995}. This corresponds to the upper limit imposed to the Mie coefficients by the energy conservation for lossless scatterers. Let us remark that resonances (thus defined) are different from the modes of the scatterer, corresponding to the poles of Mie coefficients found in the complex frequency plane, for which a scattered field may exist in the absence of an excitation field \cite{Grigoriev2013,Zambrana2015}. Expressions (\ref{an_bn_Kn}) allow us to show that this definition of resonances results in the following  condition on the K-matrix coefficients: 


\begin{eqnarray}\label{res_cond_K}
a_{n}&=1 \Rightarrow \left(K^{(e)}_{n}\right)^{-1}=0 \label{res_an_cond}, \\ 
b_{n}&=1 \Rightarrow \left(K^{(h)}_{n}\right)^{-1}=0 \label{res_bn_cond}.
\end{eqnarray}

\begin{figure}[h]
\centering
\includegraphics[width=0.5\textwidth]{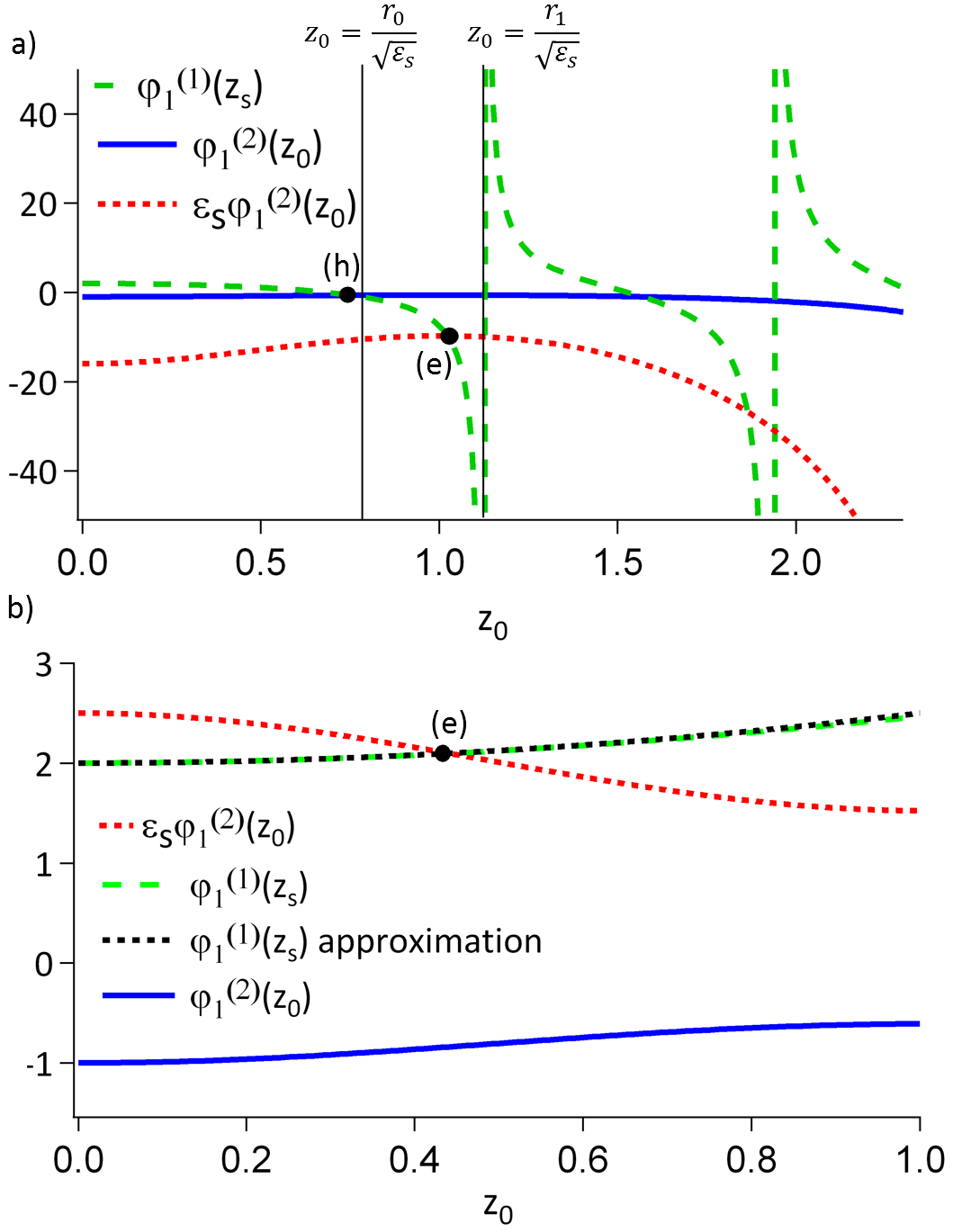}
\caption{Graphic representation of the resonance condition as a function of the size parameter $z_0$ for (a) $\varepsilon=16$  and (b) $\varepsilon=-2.5$: the electric and magnetic resonances are marked by the black dots, predictions of those resonances provided by Eqs.~\ref{res_zer_an} (full black vertical lines), function $\varphi^{(1)}_{1}(z_{s})$ (dashed green line), function $\varphi^{(2)}_{1}(z_{0})$ (full blue line), $\varepsilon_{s}\varphi^{(2)}_{1}(z_{0})$ (dotted red line). Dashed black line in (b):  $\varphi^{(1)}_{1}(z_{s})=2-\frac{z^{2}}{5}$ calculated with Eq.\label{phi_A3} at the 1st order with n=1.} \label{Fig_Res_cond_Kn} 
\end{figure}

Resonances thus correspond to the poles of the K-matrix coefficients. 
The resonance conditions provided by Eqs. (\ref{res_cond_K}) are displayed graphically for a constant and positive permittivity equal to 16 in Fig.~\ref{Fig_Res_cond_Kn}a. According to Eqs. (\ref{K_n}) and (\ref{res_cond_K}), resonances of the magnetic dipole occur at the intersections between $\varphi^{(2)}_{1}(z_{0})$ (solid blue line) and $\varphi^{(1)}_{1}(z_{s})$ (dashed green line) denoted by (h) whereas resonances of the electric dipole correspond to the intersections between $\varepsilon_{s}\varphi^{(2)}_{1}(z_{0})$ (dotted red line) and $\varphi^{(1)}_{1}(z_{s})$ (dashed green line) denoted by (e) in Fig. \ref{Fig_Res_cond_Kn}.

One can also choose to set a permittivity negative and purely real. Even if materials with such a permittivity do not exist, this can be enlightening to study what happens in this case to provide a better understanding of the plasmonic resonances. As illustrated in Fig. \ref{Fig_Res_cond_Kn}b, a resonance of the electric dipole also occurs at the intersection between $\varepsilon_{s}\varphi^{(2)}_{1}(z_{0})$ (dotted red line) and $\varphi^{(1)}_{1}(z_{s})$ (dotted green line) denoted by (e). As $\varphi^{(2)}_{1}(z_{0})$ is negative for small values of $z_{0}$, no resonance of the magnetic dipole occurs.
Although the plots of Fig. \ref{Fig_Res_cond_Kn}  only show the first solutions of the conditions (\ref{res_cond_K}) for $n=1$, one has to keep in mind that these conditions are transcendental equations and have an infinity of solutions. However, in what follows, we will mainly be interested in the first resonance of each multipolar order for subwavelength-sized scatterers. That is why we will first restrict our study to the limit $z_{0} \rightarrow 0$. In this limit, it is possible to simplify the resonance conditions by approximating the functions of $z_{0}$ by the first term of their power-series expansion :
$j_{n}(z_{0})\simeq\frac{z_{0}^{n}}{(2n+1)!!}$, $\varphi^{(1)}_{n}(z_{0})\simeq n+1$, $y_{n}(z_{0})\simeq-\frac{(2n-1)!!}{z_{0}^{n+1}}$ and $\varphi^{(2)}_{n}(z_{0})\simeq -n$, where $!!$ is defined in Appendix B. That leads to the following approximate expressions of the K-matrix coefficients:

\begin{eqnarray}\label{approx_Kne_h}
K^{(e)}_{n}&\simeq -\kappa_{n}\frac{(n+1)\varepsilon_{s}-\varphi^{(1)}_{n}(z_{s})}{n\varepsilon_{s}+\varphi^{(1)}_{n}(z_{s})} \label{appr_an_1st order},\\
K^{(h)}_{n}&\simeq-\kappa_{n}\frac{(n+1)-\varphi^{(1)}_{n}(z_{s})}{n+\varphi^{(1)}_{n}(z_{s})},
\label{appr_bn_1st order}
\end{eqnarray}

with $\kappa_{n}=\frac{z_{0}^{2n+1}}{(2n-1)!!(2n+1)!!}$. The exact expression of $\varphi^{(1)}_{n}(z_{s})$ is kept because the in-medium size parameter $z_{s}=\sqrt{\varepsilon_{s}}z_0$ is not assumed to be necessarily small.  In fact, one should keep in mind that morphological resonances, for small particles occur for large permittivity so that $z_s$ may not be small \cite{GarciaEtxarri11,Kuznetsov12}. It is then straightforward from Eqs. (\ref{res_cond_K}) and (\ref{approx_Kne_h}) to determine an approximation of the resonance conditions (\ref{res_cond_K}):

\begin{eqnarray}\label{res_zer_an}
a_{n}&=1 \Rightarrow \varphi^{(1)}_{n}(z_{s})\simeq-n\varepsilon_{s}, \label{res_an} \\ 
b_{n}&=1 \Rightarrow \varphi^{(1)}_{n}(z_{s})\simeq-n. \label{res_bn}
\end{eqnarray}

\subsection{Resonances of plasmonic scatterers:}
Only the assumption $z_0 << 1$ has been made so far but  no assumption was made about $z_s$, Eq.~\ref{res_zer_an} is valid for both metallic and dielectric particles. Let us derive the Taylor expansion to the 6$^{th}$ order of the $\varphi^{(1)}_{n}$ function \cite{Colom2016}:

\begin{equation}\label{phi_A3}
\varphi _{n}^{(T1)}(z) =n+1-\frac{z^{2}}{2n+3}-\frac{z^{4}}{(2n+5)(2n+3)^{2}}+O(z^{6}).
\end{equation}
In the limit ($z_s \rightarrow 0$), it is sufficient to consider the first term of this expansion (n+1) and it can easily be shown that Eq. (\ref{res_an}) tends towards the well-known quasi-static resonance conditions for very small plasmonic particles for electric multipoles\cite{Bohren1983}:
\begin{equation}\label{phi_A3}
\varepsilon\simeq -\frac{n+1}{n}
\end{equation}
In this same limit Eq. (\ref{res_bn}) has no solution, confirming that sub-wavelength plasmonic particles do not support magnetic resonances.
 
 \subsection{Morphological Resonances of Mie scatterers:}\label{morph_resonances}
Electric and magnetic morphological resonances in small dielectric particles can only occur when $z_{s} > 1$ requiring the permittivity to be sufficiently large \cite{Kuznetsov12}. Thus, approximations made with the assumption $z_{s} << 1$ will fail to predict the morphological resonances. That is why approximations of the Mie coefficients based on Taylor series expansion do not predict morphological resonances unless a lot of terms are taken into account. This result can be observed in Figs. \ref{Re_a1_Si_exact_Moroz} and \ref{Fig_Res_cond_Kn}  and it will be further illustrated in section \ref{sec_approx_an_bn}.\\ 
In fact, the electric morphological resonances can be better understood by studying the limit $|\varepsilon_{s}| \rightarrow \infty$. One can easily deduce from Eq. \ref{res_an} that multipolar electric resonances occur at the poles of the $\varphi^{(1)}_{n}$ functions in this limit \cite{Newton1966}. These poles correspond to the zeros of the Bessel functions \cite{Watson1944,Abramowitz1964} and in what follows, the first zero of the n-th order Bessel functions will be noted $r_{n}$. For high index dielectric scatterers for which $|\varepsilon_{s}|$ is large but not infinite, it can then be safely inferred that the first resonance of the n-th order electric multipole occurs close to the position:
\begin{equation}\label{elec_res_cond}
z_{s}\simeq r_{n}
\end{equation}
This result can be observed in Fig.~\ref{Fig_Res_cond_Kn} where the electric resonance condition is seen to be close to the pole of $\varphi^{(1)}_{1}$ in the case of $n=1$.  The exact values of $r_{0}$, $r_{1}$ and $r_{n}$ are provided in table 1 but it may be recalled that a good approximation of the l-th zero of the n-th order Bessel function can be provided by $r_{n,l} \simeq l\pi+\frac{n\pi}{2}$ \cite{Watson1944,Abramowitz1964}. At this point, one should emphasize that the condition $z_{s} = r_{n}$ actually corresponds to the first TE modes of the n-th multipole of a spherical hollow resonator (a spherical cavity in a perfect conductor) \cite{Balanis2012,Chern2010,Landau1995}. This provides some insights on the origin of morphological resonances as will be further discussed in section \ref{morphological_resonances_an_bn}.\\
A prediction of the magnetic resonance condition can then be easily deduced from Eq. \ref{res_bn} by noticing that $\varphi_n(r_{n-1}) = - n$ (see Appendix A)\cite{Newton1966}. If $|\varepsilon_{s}|$ is large, it can then be assumed that the first resonance of the n-th magnetic multipole occurs close to the position:
\begin{equation} \label{magnetic_res_cond}
 z_{s}\simeq r_{n-1} 
\end{equation}
This is confirmed in Fig.~\ref{Fig_Res_cond_Kn} in the case of $n=1$ where one clearly sees that the magnetic resonance is close to this condition. This resonance condition differs from the first TM mode of a spherical hollow resonator occurring for $ z_{s}\simeq r^{'}_{n} $, $r^{'}_{n}$ being the first zero of the derivative of the n-th order Bessel functions \cite{Balanis2012,Landau1995}. This will be further discussed in section \ref{morphological_resonances_an_bn}.\\ 
In order to get a good approximation of $b_{n}$ at the vicinity of the magnetic resonance, one could then choose to approximate $\varphi _{n}^{(1)}(z_{s})$ by its power series expansion around $z_{s}=r_{n-1}$ (calculations are made in Appendix D):
\begin{eqnarray}\label{phi_A4}
\varphi _{n}^{(T2)}(z) \simeq -n- r_{n-1}\left(z - r_{n-1} \right) \cr
- \left(n+1\right)\left(z - r_{n-1} \right)^2 \cr 
-\frac{1}{3}\left(\frac{n(2n+1)}{r_{n-1}}\right)\left(z - r_{n-1} \right)^3
\end{eqnarray}
Fig.~\ref{Fig_Taylor} shows that Eq.~\ref{phi_A4} provides a very good approximation of $\varphi _{n}^{(1)}(z_{s})$, but only on a small interval of size parameters close to the resonance.\\ 
It can then be concluded from this study that the slow convergence of the Taylor series expansions does not allow accurate and compact approximated expressions of the $\varphi_n^{(1)}$ functions. It will be confirmed by the results obtained on section \ref{sec_approx_an_bn}.

\section{Weierstrass approximations of $\varphi _{n}^{(1)}$}\label{appr_phi_n}

\begin{figure}[h]
\centering
\includegraphics[width=0.45\textwidth]{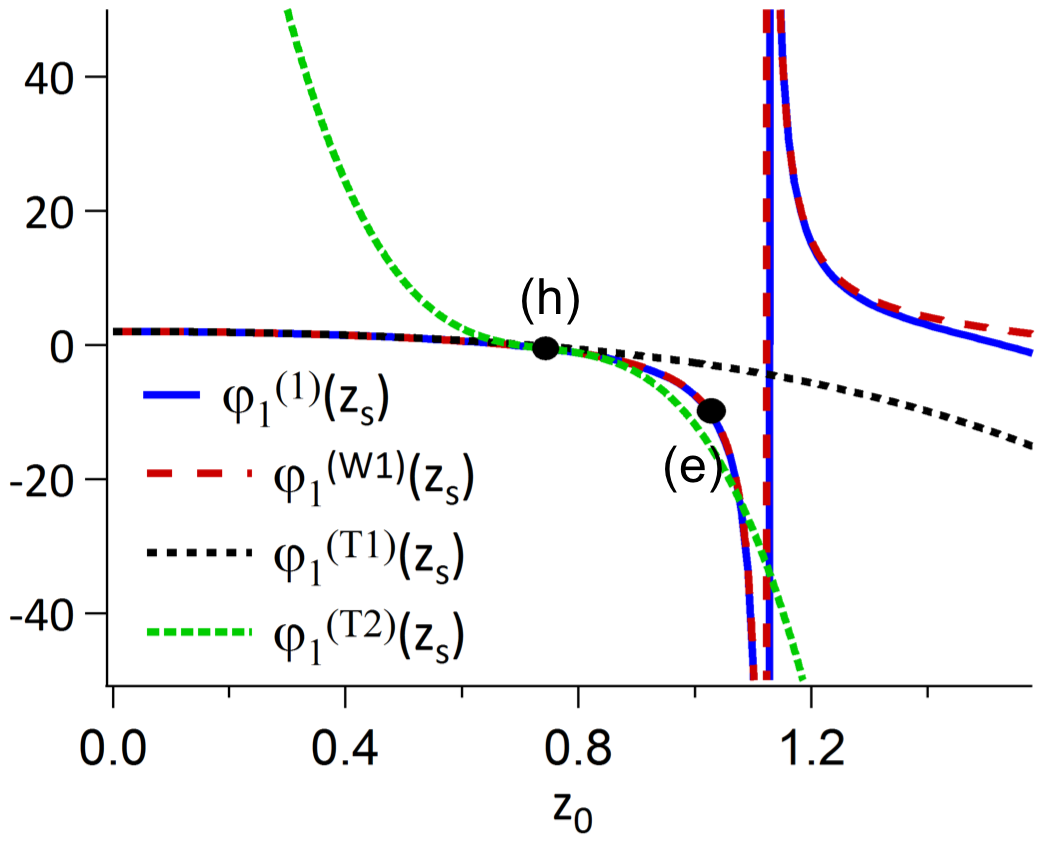}
\caption{Comparison of the approximations of the $\varphi^{(1)}_{1}(z_{s})$ function. Exact calculation (full blue line). $\varphi_{1}^{(T1)}(z_{s})$: Taylor expansions around $z_{s}=0$ (dashed black line),$\varphi_{1}^{(W1)}(z_{s})$: approximation derived in Eq. \ref{phi_A1} (dashed red line), $\varphi_{1}^{(T2)}(z_{s})$: Taylor expansion  around  $z_{s}=r_{n-1}$ derived in Eq. \ref{phi_A4} (dotted green line). } \label{Fig_Taylor} 

\end{figure}

We propose to address this issue of the slow convergence of the Taylor descriptions in the proximity of the poles of $\varphi _{n}^{(1)}(z_{s})$ by using the Weierstrass expansion of the Bessel function \cite{Watson1944,Abramowitz1964}: 

\begin{equation}
j_{n}(z)=\frac{z^{n}}{(2n+1)!!}\prod\limits_{l=1}^{\infty }\left(1-\left(\frac{z}{%
r_{n,l}}\right)^{2}\right),
\label{weier_fact_j}
\end{equation}
where $r_{n,l}$ is the l-th zero of the n-th order Bessel function.
The expression of the $\varphi _{n}^{(1)}$ function can then be deduced from Eqs. (\ref{express_phi_n}) and (\ref{weier_fact_j}) (see Appendix B)\cite{Grigoriev2015,Devilez2015}:
\begin{equation}\label{Weir_fact_phi_n_(1)}
\varphi _{n}^{(1)}(z)=n+1+\sum\limits_{l=1}^{\infty }\frac{2z^{2}}{%
z^{2}-(r_{n,l})^{2}}.
\end{equation} 

Expression (\ref{Weir_fact_phi_n_(1)}) is an exact expansion of $\varphi _{n}^{(1)}$ which takes into account the existence of an infinite number of poles located on the real axis, as observed in Fig.\ref{Fig_Res_cond_Kn}, and corresponding to the zeros of $j_{n}$. It is then interesting to recall the asymptotic form of these zeros for large values of l: $r_{n,l} \simeq l\pi+ n\pi/2$ \cite{Watson1944,Abramowitz1964}. In the previous section, it was shown that those poles are of great importance in the emergence of the electric morphological resonances and that is why it is necessary to find approximations of $\varphi _{n}^{(1)}$ featuring the same poles. In our study, we are seeking for approximations capable to predict the first morphological resonance. Approximations of $\varphi _{n}^{(1)}$ can be obtained by truncating the infinite sum in Eq.~(\ref{Weir_fact_phi_n_(1)}) and by conserving only its first term. But, rather than completely neglecting the influence of higher order poles, one can also approximate their contributions. As shown in Appendix \ref{App_approx_phi_n}, if we consider that $z_{s} <<  r_{n,2}$, we obtain the following approximation: 
\begin{equation}\label{phi_A1}
\varphi _{n}^{(W1)}(z_0) = n + 1+\frac{2z_n^{2}}{z_n^{2}-1}+ 2\rho^{(e)}_n z_0^2,
\end{equation}
where we set for compact notations that $z_n \equiv z_0/r_{n,1} \equiv z_0/r_{n}$ and $\rho^{(e)}_n \equiv \frac{1}{r^2_{n}} - \frac{1}{2(2n+3)}$, with $r_{n}$ being the first zero of $j_{n}$. 
Regarding the approximation of the magnetic coefficients, we have seen in section \ref{section_res_cond} that their first resonance occurs near the condition $z_{s}= r_{n-1}$. In order to have a good prediction of the magnetic resonances, an accurate approximation of $\varphi _{n}^{(1)}(z_{s})$ near $z_{s}= r_{n-1}$ must be found. As seen in the previous section and in Fig. \ref{Fig_Taylor}, a simple power series expansion of $\varphi _{n}^{(1)}(z_{s})$ does not provide satisfying results. A better approximation has been found under the following form:
\begin{equation}\label{phi_A2}
\varphi _{n}^{(W2)}(z_0) = n + 1+\frac{2z_n^{2}}{z_n^{2}-1}+ 2\rho^{(h)}_n z_0^2,
\end{equation}
where $\rho^{(h)}_n$ has been derived to impose $\varphi _{n}^{(A2)}(z)=-n$. It can then be easily shown that $\rho^{(h)}_n\equiv \frac{1}{r_{n}^2-r_{n-1}^2}-\frac{2n+1}{2r_{n-1}^2}$. 

Approximations of the Bessel functions can also be derived by following a similar approach leading to the subsequent expression (see Appendix \ref{App_approx_phi_n}):
\begin{equation}\label{j_A1}
j^{(W1)}_{n}(z_0)=\frac{z_0^{n}}{(2n+1)!!}\left(1-z_n^{2}\right)e^{\rho_n z_0^{2}}.
\end{equation}

\begin{table}
\caption{Numerical values  of the constants employed in the article for the first multipole orders.}
\begin{ruledtabular}
\begin{tabular}{cccc}
  x & $r_n$ & $\rho_n^{(e)}$ & $\rho_n^{(h)}$ \\
 \hline
$n=0$ & $\pi$ & -0.065 & x \\ 
 \hline 
$n=1$ & 4.49 & -0.05 & -0.055\\ 
 \hline
$n=2$ & 5.76 & -0.041 & -0.047\\
 \hline
\end{tabular}
\end{ruledtabular}
\end{table}
The approximations obtained for the special functions appearing in the Mie theory can now be used to find approximations of the Mie coefficients. 

\section{Approximations of $a_{n}$ and $b_{n}$}\label{sec_approx_an_bn}
In order to find an accurate approximation of the $a_{n}$ and $b_n$ coefficients, in particular at the vicinity of their resonances \cite{Garcia-Camara2008,Videen1992,Miroshnichenko2009}, we start from the exact expression (\ref{Mie_a}) and make use of the approximations (\ref{phi_A1}) and (\ref{j_A1}) derived with the sole assumption $z_{s} << r_{n,2}$. If the exact expressions of $h_{n}^{(+)}$ and $\varphi _{n}^{(+)}$ are kept, it can be shown, provided several steps of calculations, that the Mie coefficients can be cast (see Appendix \ref{AppE} and Appendix \ref{AppF}):

\begin{eqnarray}
\begin{aligned}
a^{(A1)}_{n} =\frac{(n+1)z^{2n+1}}{(2n+1)!!}\frac{e^{-iz+\rho^{(e)}_{n}z^{2}}}{Q_{n}(z)}\times &\cr
\frac{(\varepsilon-1)\left(f_{n}(\varepsilon,z)-z_{n}^{2}\right)}{\varepsilon g_{n}(z)f_{n}(\varepsilon,z)-(n+1)} \label{appr_an_A1}\\
b^{(A1)}_{n}=\frac{z^{2n+1}}{(2n+1)!!}\frac{e^{-iz+\rho_{n}^{(h)}z^{2}}}{Q_{n}(z)}\times &\cr 
\frac{(\varepsilon-1)L_{n}(\varepsilon,z)}{\varepsilon L_{n}(\varepsilon,z)-(n+1)+\varphi_{n}^{(+)}(z)}\label{appr_bn_A1}
\end{aligned}
\end{eqnarray}

with $g_{n}(z)=\varphi_{n}^{(+)}(z)-2\rho_{n}z^{2}$ and $Q_{n}(z)$ is a polynomial function detailed in Appendix \ref{AppG} and $\varphi_{n}^{(+)}(z)$ being simply calculated thanks to equations (\ref{express_phi_n}) and (\ref{hn_plus}). In the electric coefficient expression, $f_{n}(\varepsilon,z)=\frac{1-\varepsilon z_n^{2}}{1-\frac{n+3}{n+1}\varepsilon z_n^{2}}$ while in the magnetic coefficient expression $L_{n}(\varepsilon,z)=-\frac{2z_n^{2}}{\varepsilon z_n^{2}-1}-2\rho_{n}^{(h)}z^{2}$.

Comparisons between these approximations and exact calculations are shown in the following figures. In order to further highlight the relevance of our study, we also make comparisons with approximations already derived in literature \cite{Moroz2009,Colom2016} that are based on power series expansions of the $K_{n}$ coefficients in Eqs. \ref{an_Kn_e} and \ref{an_Kn_h}.

\begin{eqnarray}
\begin{aligned}
\left(a_{1}^{(T1)}\right)^{-1} &=-i\left(-\frac{3(\varepsilon + 2)}{2z^{3}(\varepsilon -1)}+\frac{9(\varepsilon - 2)}{10z(\varepsilon -1)}\right)+1\cr
\left(b_{1}^{(T1)}\right)^{-1}&=i\frac{45}{z^{5}(\varepsilon -1)}-\frac{15i(2\varepsilon - 5)}{7z^{3}(\varepsilon -1)}\\
&-\frac{i(\varepsilon^{2} +100\varepsilon - 125)}{49z(\varepsilon -1)}+\\
\left(a_{1}^{(T2)}\right)^{-1}&=i\frac{3(\varepsilon + 2)}{2z^{3}(\varepsilon -1)}-\frac{9i(\varepsilon - 2)}{10z(\varepsilon -1)}\\
&-\frac{9iz(\varepsilon^{2} - 24\varepsilon + 16)}{700(\varepsilon -1)}+1\cr
\end{aligned}\label{approximations_a1_A3_A4}
\end{eqnarray}

\begin{figure}[h]
\centering
\includegraphics[width=0.45\textwidth]{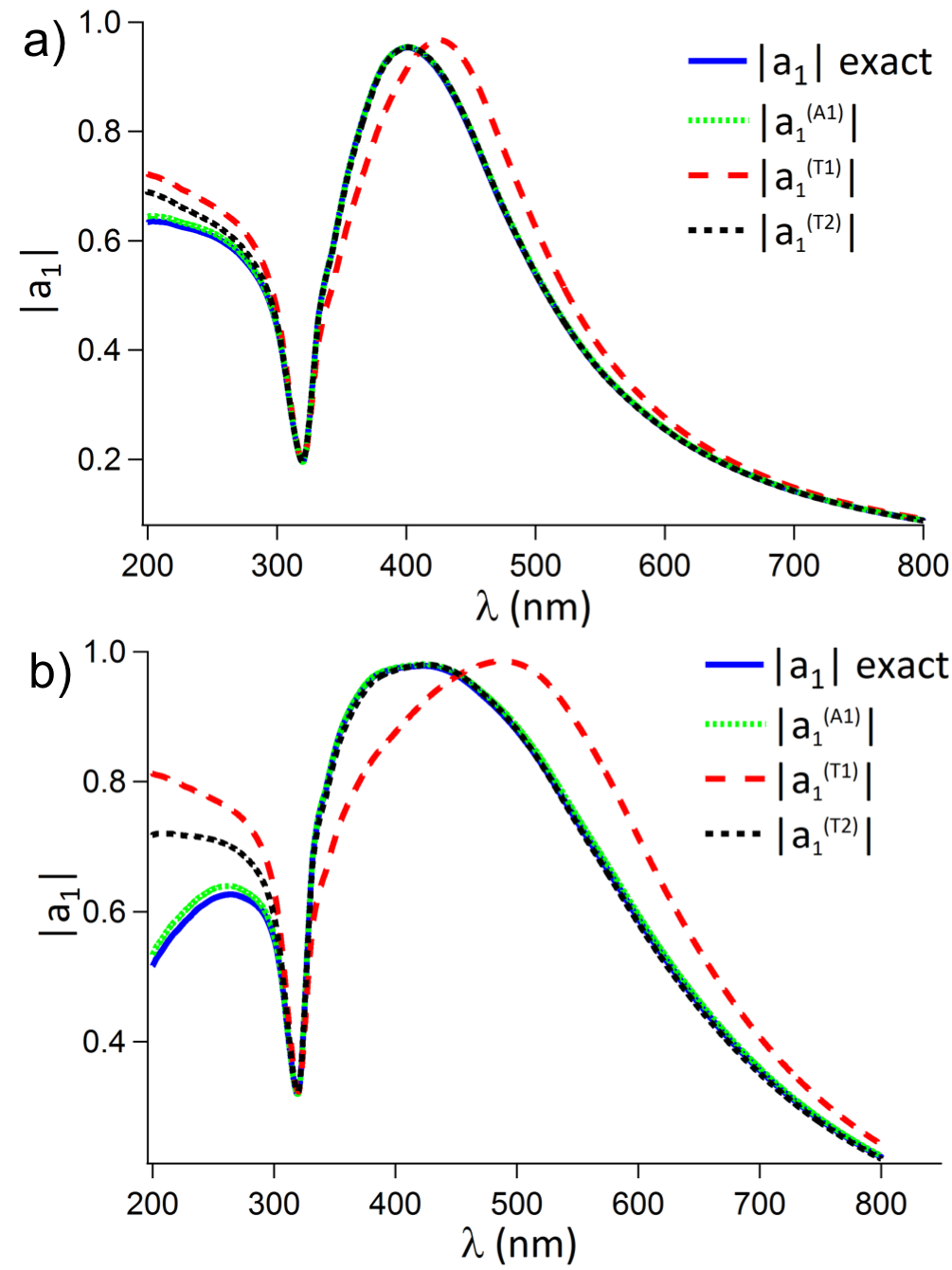}
\caption{Comparison between exact (full blue line) and approximations (\ref{appr_an_A1}) (dotted green line) and (\ref{approximations_a1_A3_A4}) (dashed red and black lines) of $a_{1}$ for a sphere of silver \cite{Johnson} 60 nm (a) and 80 nm (b) in radius.} \label{a1_exact_approx_Silver} 
\end{figure}

\begin{figure}[h]
\centering
\includegraphics[width=0.45\textwidth]{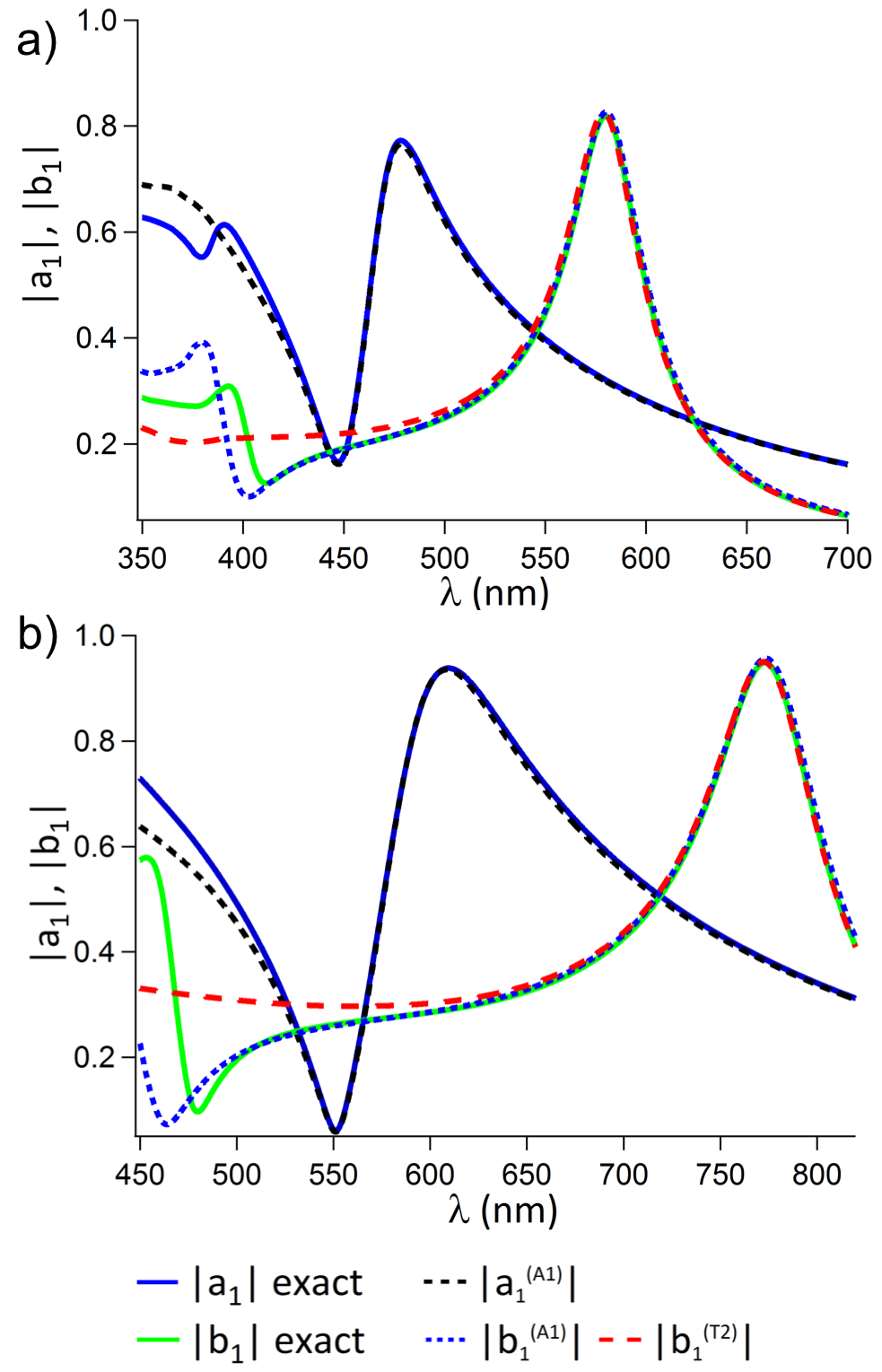}
\caption{Comparison between exact calculations of $a_{1}$ (full blue line) and $b_{1}$ (full green line) with approximations (\ref{appr_an_A1}) of $a_{1}$ (dashed black line) and $b_{1}$ (dotted blue line) and power-series approximations of $b_{1}$ (\ref{approximations_a1_A3_A4}) (dashed red line), for a sphere of silicon \cite{Aspnes1983} of radius 70 nm (a) and 100 nm (b).} \label{a1_b1_exact_approx_Si} 
\end{figure}

\begin{figure}[h]
\centering
\includegraphics[width=0.5\textwidth]{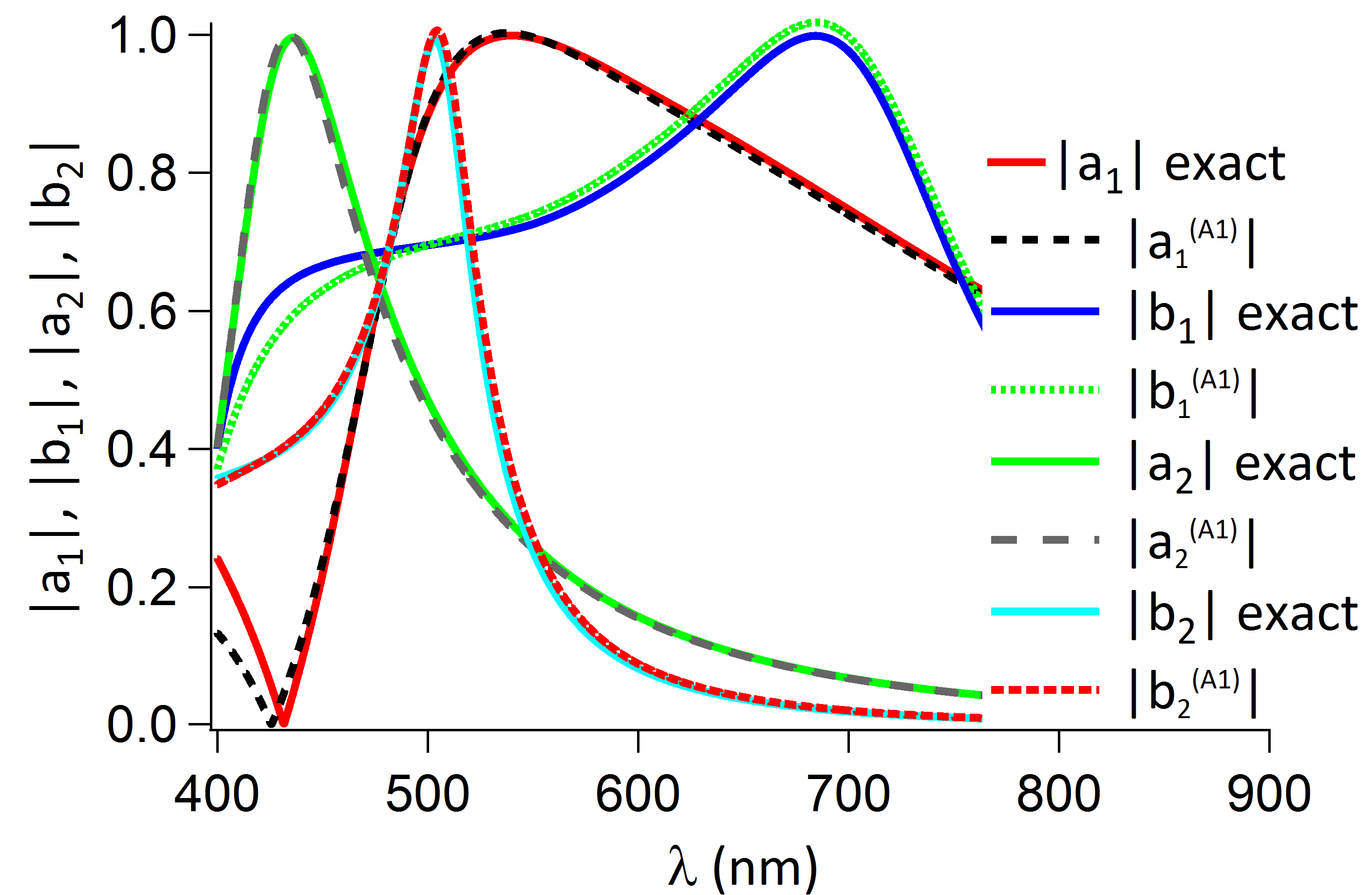}
\caption{Comparison between exact calculations and approximations \ref{appr_an_A1} for:  $a_{1}$ (full red line and dashed black line), $b_{1}$ (full blue line and dotted green line), $a_{2}$ (full green line and dashed gray line) and $b_{2}$ (full cyan line and dotted red line) for a $TiO_{2}$ \cite{Yamada1999} 140 nm in radius.} \label{a1_b1_a2_b2_exact_approx_TiO2} 
\end{figure}

It is clearly observed in Fig. \ref{a1_exact_approx_Silver} that our approximations achieve to reproduce the resonances predicted by exact calculations in a more accurate way than state-of-the-art approximations, even quite lengthy high-order Taylor expansions.
Although we did not explicitly derive these approximations for describing plasmonic scatterers, one clearly sees that these new approximations are more accurate than the approximations (\ref{approximations_a1_A3_A4}) as can be seen in Fig.\ref{a1_exact_approx_Silver}.\\
However, the main interest of these new approximations is that they are highly accurate for high-index dielectric scatterers as shown in Fig. \ref{a1_b1_exact_approx_Si} for a silicon scatterer. 
In fact, it was already shown in Fig. \ref{Re_a1_Si_exact_Moroz} that no approximation based on Taylor series expansions achieves to predict the resonance of $a_{1}$ but the approximation derived in this study (\ref{appr_an_A1}) does predict these resonances accurately. Regarding the magnetic resonances, even though the approximation  $b_{1}^{(T1)}$ in Eq. (\ref{approximations_a1_A3_A4}) which is a high-order power-series expansion of $b_{1}$ shows the dipolar magnetic resonance \cite{Colom2016}, our approximation stays more accurate for a larger range of sizes and wavelengths.\\
We now aim at studying the validity of these expressions in the case of larger particles made of lower refractive index. This will allow us to test the accuracy of higher orders expressions, in particular quadrupolar orders. For that purpose, we consider a sphere made of  $TiO_{2}$, 140 nm in radius. We compare the calculations of the dipolar and quadrupolar electric and magnetic Mie coeffcients obtained by exact calculations (Eq. \ref{Mie_a}) with the expressions (\ref{appr_an_A1}). The plot in Fig. \ref{a1_b1_a2_b2_exact_approx_TiO2} shows the very good accuracy of these expressions for dipolar and quadrupolar orders, even when considering larger particles made of lower refractives indices.

\section{Discussion}\label{morphological_resonances_an_bn}
\begin{figure}[h]
\centering
\includegraphics[width=0.5\textwidth]{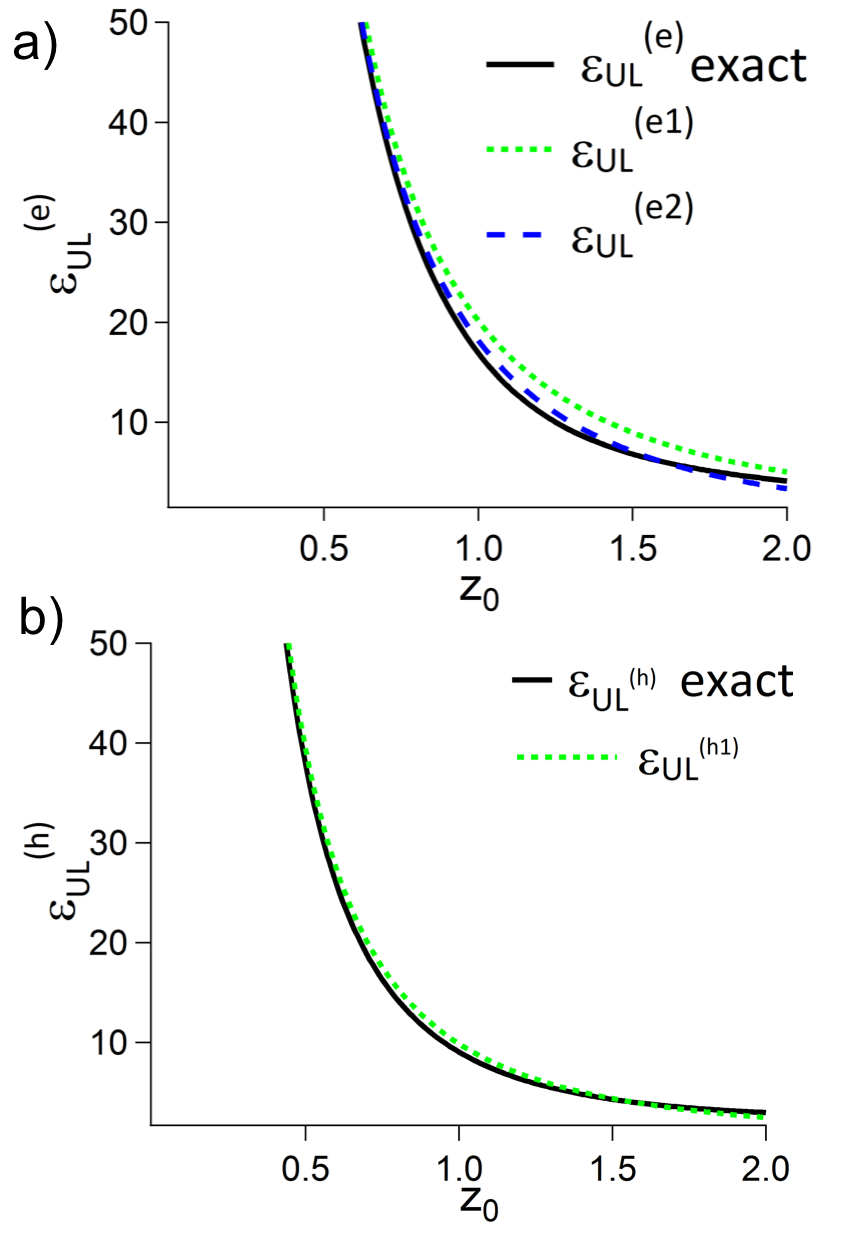}
\caption{Comparison between the exact $\varepsilon_{UL}$ (black full line)required to reach the resonance and the prediction provided by Eqs. (\ref{elec_res_cond}) and (\ref{magnetic_res_cond}) respectively labelled $\varepsilon_{UL}^{(e1)}$ and $\varepsilon_{UL}^{(h1)}$ (dotted green line). A better approximation of $\varepsilon_{UL}^{(e)}$ is obtained by solving $\varphi _{n}^{(W1)}(z_s) =-n\varepsilon_{s}$, labelled $\varepsilon_{UL}^{(e2)}$ (dashed blue line)}\label{unit_limi_e_h} 
\end{figure}

Conditions of resonance derived in section \ref{morph_resonances} have been very useful to derive accurate approximations of the Mie coefficients in the previous section. Here, we will show that they can also provide more insight on the origin of morphological resonances. Since condition (\ref{elec_res_cond}) is close to the TE mode of a hollow resonator, one can infer that morphological resonances occur thanks to the ability of high index dielectric scatterers to play the role of a cavity. Since high index dielectric scatterers are not perfect cavities, the trapped electromagnetic field leaks in the surrounding medium driving to resonances of the scattered field. When $|\varepsilon_{s}| \rightarrow \infty$, the scatterer becomes a very good cavity for the electromagnetic field which can be trapped inside the resonator for a long time. It is not surprising then to find the same resonance condition as the one of a hollow resonator in this case.\\ 
Nonetheless, the resonance condition (\ref{magnetic_res_cond}) for magnetic multpoles is different from the TM mode of a hollow resonator. This TM mode normally occurs at $z_{s}\simeq r^{'}_{n} $, $r^{'}_{n}$ being the first zero of the derivative of the n-th order Bessel functions \cite{Balanis2012} and not $z_{s}\simeq r_{n-1}$. For $n=1$,  $r^{'}_{1}$ and $r_{n-1}=r_{0}$ take the following value $r^{'}_{1}=2.744$ and $r_{0}=\pi$. However, one can infer that magnetic morphological resonances also occur thanks to the ability of high idex dielectric scatterers to concentrate light. Since even in the limit $|\varepsilon_{s}| \rightarrow \infty$, these high-index dielectric scatterers are not perfect cavities and that explains why the magnetic resonances are different from the TM modes of hollow resonators.\\
In section \ref{morph_resonances}, these resonance conditions were derived in the limit $|\varepsilon_{s}| \rightarrow \infty$. One could then question the validity of such conditions of resonance for large but not infinite values of $|\varepsilon_{s}|$. A comparison between the exact values of $|\varepsilon_{s}|$ required to reach the resonance, also called unitary limit \cite{Colom2016}, and the predictions provided in section \ref{morph_resonances} needs then to be carried out. The exact values of the permittivity needed to reach the resonance for a given $z_{0}=kR$ can be derived by numerically solving the equation $a_{1}=1$ for the electric dipole resonance and $b_{1}=1$ for the magnetic dipole resonance.
In Fig. \ref{unit_limi_e_h}, the exact value of the unitary limit permittivity for the electric dipole $\varepsilon_{UL}^{(e)}$ in Fig. \ref{unit_limi_e_h} is compared to the prediction provided by Eq. (\ref{elec_res_cond}) in section \ref{morph_resonances}: $z_{s}=r_{1}$ or equivalently: $\varepsilon_{s}^{(e1)} = \left(\frac{r_{1}}{z}\right)^{2}$. One can clearly see that this expression predicts accurately the asymptotic behavior of the exact $\varepsilon_{UL}^{(e)}$ for very small $z_{0}$ but is not very accurate for larger $z_{0}$. 
In Fig. \ref{unit_limi_e_h}, the same comparison is also carried out beween the exact unitary limit permittivity for the magnetic dipole $\varepsilon_{UL}^{(h)}$ and the prediction given by Eq. (\ref{magnetic_res_cond}), $\varepsilon_{s}^{(h1)} = \left(\frac{r_{0}}{z}\right)^{2}$. A very good agreement is observed between the exact value and the prediction.\\
A more accurate prediction of $\varepsilon_{UL}^{(e)}$ can also be derived. To do so, one could solve Eq. (\ref{res_an}) for $n=1$: $\varphi _{1}^{(1)}(z_s) =-\varepsilon_{s}$. However, this equation can only be solved numerically. On the other hand, if the approximation $\varphi _{1}^{(W1)}(z_s)$ given by Eq. (\ref{phi_A1}) is used, the previous equation reduces to a second order equation in $\varepsilon$ and can be analytically solved leading to the prediction $\varepsilon_{UL}^{(e2)}$ (the exact expression of $\varepsilon_{UL}^{(e2)}$ is provided in appendix \ref{AppH}). This prediction proves to be quite accurate for a large range of $z_{0}$ as can be seen in Fig. \ref{unit_limi_e_h}.

Finally it is also interesting to notice that the approximation (\ref{appr_an_A1}) also predicts a zero of the Mie coefficients different from the trivial condition ($\varepsilon-1$). These zeros actually correspond to the anapoles in \cite{Miroshnichenko2015}. They are in fact reached when $f_{n}(\varepsilon,z)-z_n^{2}=0$ or equivalently $\frac{1}{z^{2}_n}f_{n}(\varepsilon,z)=1$. This latter condition can be found while searching for solutions to $\varepsilon_{eq}=1$, the definition of $\varepsilon_{eq}(z)=\left(\frac{r_{1}}{z}\right)^{2}\frac{1-\varepsilon(z/r_{1})^{2}}{1-2\varepsilon(z/r_{1})^{2}}$ being provided in \cite{Devilez2015}. Expression (\ref{appr_an_A1}) also provides a condition for which $b_{n}$ is null. The non trivial solution, different from the trivial solution $\varepsilon=1$, corresponds to $L_{n}(\varepsilon,z)=0$. 

\section{Conclusion}
To conclude, the use of the $K$-matrix has allowed us to derive  resonance conditions for both plasmonic and high-index dielectric resonant particles. We have thus been able to show that under the condition $|\varepsilon_{s}|>>1$, different from the condition used for Taylor series $z_{s} << 1$, the electric resonance is close to $z_{s}\simeq r_{n}$ and the magnetic resonance is close to $z_{s}\simeq r_{n-1}$.  The proximity of the electric resonance to the pole of the $\varphi_n(z_s)$ function at $z_s= r_{n}$ explains the weak convergence of the Taylor series expansion for approximating Mie coefficients especially near the electric resonances. We proposed to solve this problem by using a Weierstrass expansion of the Bessel functions. This method allows us to derive for any multipolar order highly accurate electric and magnetic polarizability expressions. We evidenced the high accuracy of these expressions by calculating the dipolar and quadrupolar polarizability expressions of spherical particles made of silver, silicon and titania. These expressions bring novel analytical tools to explain the resonant light interaction with metallic or dielectric particles. They also permitted us to bring more physical insight on the origin of morphological resonances. In particular, these formulations allowed us to calculate a very accurate prediction of the dielectric permittivity required to reach the resonance, also called unitary limit. Such expressions will offer novel opportunities for modeling the light scattering in complex media or for homogenizing optical systems made of resonant light scatterers. 
 
\section{acknowledgments}
Research conducted within the context of the International Associated Laboratory ALPhFA: Associated Laboratory for Photonics between France and Australia. This work has been carried out thanks to the support of the A*MIDEX project (no. ANR-11-IDEX-0001-02) funded by the Investissements d'Avenir French Government program, managed by the French National Research Agency (ANR).

\appendix

\section{Resonance and zero conditions of $b_{n}$}\label{AppA}
It has been shown in section \ref{section_res_cond} that a good prediction of the first resonance of $b_{n}$ in the limit $z_{0}<<1$ could be provided by the solution of the following equation:
\begin{equation} \label{demo_bn_res}
\varphi^{(1)}_{n}(z)=-n 
\end{equation}
If we use the following recurrence relation for spherical Bessel functions (ref.\cite{Abramowitz1964},p.439):
\begin{equation}\label{recur_rel_jn}
\begin{split}
j_{n}'(z)=-\frac{n+1}{z}j_{n}(z)+j_{n-1}(z)\cr
\Rightarrow [zj_{n}(z)]'=-nj_{n}(z)+zj_{n-1}(z)\cr
\Rightarrow \varphi^{(1)}_{n}(z)=-n+z\frac{j_{n-1}(z)}{j_{n}(z)}
\end{split}
\end{equation}
It then follows that Eq. \ref{demo_bn_res} is verified for the zeros of Bessel functions of order $n-1$: $z=r_{n-1,l}$.

\section{Weierstrass factorization of Bessel functions}\label{AppB}
As demonstrated by Watson in \cite{Watson1944} (p 497-498), it is possible to express cylindrical Bessel functions as an infinite product of factors involving their zeros: 
\begin{equation}
J_{\nu }(z)=\frac{1}{\Gamma (\nu +1)}\left( \frac{z}{2}\right) ^{\nu
}\prod\limits_{n=1}^{\infty }\left( 1-\left( \frac{z}{z_{\nu ,n}}\right)
^{2}\right)\label{weier_fact_Jn}
\end{equation}
where $z_{\nu ,n}$ is the $n^{th}$ zero of the cylindrical Bessel functions of order $\nu$. This expression can be generalized to the spherical Bessel functions $j_{n}$ by means of the relation: $j_{n}(z)=\sqrt{\frac{\pi }{2z}}J_{n+1/2}(z)$. If we set $r_{n,j}\equiv z_{n+1/2,j}$ and if we notice that $\Gamma (n+1/2)=2^{-n}\sqrt{\pi}(2n-1)!!$, we get the following expression (\ref{weier_fact_j}): 
\begin{equation}
j_{n}(z)=\frac{z^{n}}{(2n+1)!!}\prod\limits_{l=1}^{\infty }\left(1-\left(\frac{z}{%
r_{n,l}}\right)^{2}\right)
\end{equation}
The double factorial operator $!!$ is defined such that:
\begin{equation}
n!!=\prod_{k=0}^{m} \left(n-2k\right) = n(n-2)(n-4)\ldots
\end{equation}
where $m={\rm Int}\left[ (n+1)/2 \right]-1 $ with $0!!=1$; or in terms of ordinary factorials $via$ the
relations $(2n-1)!!=\frac{(2n)!}{2^n n!}$ and $(2n)!!=2^n n!$ for $n=0,1,2,\ldots $.

These expressions are designated as Weierstrass factorizations throughout the article as it can also be obtained by using the Weierstrass factorization theorem.
It can then be used to derive an expression of $\varphi_{n}^{(1)}$ functions also appearing in our formulation of the Mie coefficients. One should first notice that $\varphi_{n}^{(1)}$ as defined in (\ref{express_phi_n}) is equal to $z$ times the logarithmic derivative of the Ricatti Bessel functions $zj_{n}(z)$. From (\ref{weier_fact_j}), it is then straightforward to show that:
\begin{eqnarray}
\begin{aligned}
&\lefteqn{\varphi_{n}^{(1)}(z) =} \cr
&z\left( (n+1)\frac{1}{z}+\sum\limits_{l=1}^{\infty}\left( -\frac{2z}{(r_{n,l})^{2}}\right) \frac{1}{1-\left( \frac{z}{r_{n,l}}\right) ^{2}}\right),\\  
&\varphi_{n}^{(1)}(z) =n+1+\sum\limits_{l=1}^{\infty }\frac{2z^{2}}{z^{2}-(r_{n,l})^{2}}.
\end{aligned}
\end{eqnarray} 

\section{Approximation of $\varphi _{n}^{(1)}$ for $a_{n}$}\label{App_approx_phi_n}
As suggested in \cite{Grigoriev2013} (see notably the supplementary material), the expressions derived in Appendix \ref{AppA} can be used to approximate functions $j_{n}$ and $\varphi_{n}^{(1)}$ as an alternative to their Taylor series expansions. It was this method that was followed to derive the approximations (\ref{phi_A1}), (\ref{phi_A2}) and (\ref{j_A1}). Here, we provide a demonstration of these two expressions. $\varphi_{n}^{(1)}$ is equal to:
\begin{equation}
\begin{aligned}
\varphi _{n}^{(1)}(z) &=n+1+\frac{2z^{2}}{z^{2}-r_{n}^{2}}%
+\sum\limits_{l=2}^{\infty }\frac{2z^{2}}{z^{2}-(r_{n,l})^{2}} \cr
&=n+1+\frac{2z^{2}}{z^{2}-r_{n}^{2}}-2z^{2}\sum\limits_{l=2}^{\infty }%
\frac{1}{(r_{n,l})^{2}}\frac{1}{1-\frac{z^{2}}{(r_{n,l})^{2}}}
\end{aligned}
\end{equation}
As in our study $z<<r_{n,2}$, for $l\geq2$ $\frac{1}{1-\frac{z^{2}}{(r_{n,l})^{2}}}\simeq1$ which leads to: 
\begin{equation}
\varphi _{1}^{(1)}(z) \simeq2+\frac{2z^{2}}{z^{2}-r_{n}^{2}}-2z^{2}\sum\limits_{l=2}^{\infty }\frac{1}{(r_{1,l})^{2}}.\label{Ml_phi1}  
\end{equation}
Finally, as $\sum\limits_{l=1}^{\infty }\frac{1}{(r_{n,l})^{2}}=\frac{1}{2(2n+3)}$ (see \cite{Watson1944}, p 502), Eq.~\ref{Ml_phi1} simplifies to:
\begin{equation}\label{dem_phi_A2}
\varphi _{n}(z) \approx n + 1+\frac{2z^{2}}{z^{2}-(r_n)^{2}}+ 2\rho_n z^2,
\end{equation}
where $\rho_n \equiv \frac{1}{r^2_{n}} - \frac{1}{2(2n+3)}$.

Moreover, we can apply the same idea to approximate the spherical bessel functions $j_{n}$:\begin{equation}
j_{n}(z)=\frac{z^{n}}{(2n+1)!!}\left(1-\left(\frac{z}{r_{n,1}}\right)^{2}\right)\prod\limits_{l=2}^{\infty }\left(1-\left(\frac{z}{r_{n,l}}\right)^{2}\right),
\end{equation}  
and $\prod\limits_{l=2}^{\infty }\left(1-\left(\frac{z}{r_{n,l}}\right)^{2}\right)$ can be approximated: 
\begin{equation}
\begin{aligned}
\prod\limits_{l=2}^{\infty }\left(1-\left(\frac{z}{r_{n,l}}\right)^{2}\right)&=exp(ln(\prod\limits_{l=2}^{\infty }\left(1-\left(\frac{z}{r_{n,l}}\right)^{2}\right))) \cr
&=exp(\sum\limits_{l=2}^{\infty }ln\left(1-\left(\frac{z}{r_{n,l}}\right)^{2}\right)).
\end{aligned}
\end{equation}
If $z<<r_{n,2}$, $\sum\limits_{l=2}^{\infty}ln\left(1-\left(\frac{z}{r_{n,l}}\right)^{2}\right) \to -\sum\limits_{l=2}^{\infty}\left(\frac{z}{r_{n,l}}\right)^{2}$, it then follows from the previous results that:
\begin{equation}
\prod\limits_{l=2}^{\infty }\left(1-\left(\frac{z}{r_{n,l}}\right)^{2}\right)\simeq exp(\rho_n z^{2}),
\end{equation}
which leads to the following approximation for $j_{n}$:
\begin{equation}
j_{n}(z)\simeq\frac{z^{n}}{(2n+1)!!}\left(1-\left(\frac{z}{r_{n,1}}\right)^{2}\right)exp(\rho_n z^{2})
\end{equation} 

\section{Approximation of $\varphi _{n}^{(1)}$ for $b_{n}$}
In order to approximate the function $\varphi _{n}^{(1)}$ at the vicinity of the resonance condition of $b_{n}$, we can make the choice to take the power series expansion around $z_{s}=r_{n-1}$:
\begin{equation}
\varphi _{n}^{(1)}(z)\simeq\varphi _{n}^{(1)}(r_{n-1})+(z-r_{n-1})\left.\frac{d\varphi _{n}^{(1)}}{dz}\right\vert_{r_{n-1}}+...
\end{equation}
\begin{equation}
\begin{aligned}
\frac{d\varphi _{n}^{(1)}}{dz}(z)&=\frac{d}{dz}\left(z\frac{j_{n-1}(z)}{j_{n}(z)}\right)\cr
&=\frac{j_{n-1}(z)}{j_{n}(z)}+zj_{n-1}(z)\frac{d}{dz}\left(\frac{1}{j_{n}(z)}\right)\cr
&+\frac{z}{j_{n}(z)}\frac{dj_{n-1}}{dz}(z)
\end{aligned}
\end{equation} 
which leads to:
\begin{equation}
\left.\frac{d\varphi _{n}^{(1)}}{dz}\right\vert_{r_{n-1}}=\frac{r_{n-1}}{j_{n}(r_{n-1})}j'_{n-1}(r_{n-1})
\end{equation}
By using a recurrence relation for spherical Bessel functions (ref.\cite{Abramowitz1964},p.439):
\begin{equation}\label{recur_rel_jn2}
-\frac{n}{z}j_{n}(z)+j_{n}'(z)=-j_{n+1}(z),
\end{equation}
we can show that $j'_{n-1}(r_{n-1})=-j_{n}(r_{n-1})$. This result leads to:
\begin{equation}
\left.\frac{d\varphi _{n}^{(1)}}{dz}\right\vert_{r_{n-1}}=-r_{n-1}
\end{equation}  
Similar calculations allow us to show that $\varphi _{n}^{(1)''}(r_{n-1})=-2(n+1)$ and $\left.\frac{d^3\varphi _{n}^{(1)}}{dz^3}\right\vert_{r_{n-1}}=-2\frac{n(2n+1)}{r_{n-1}}-2r_{n-1}$.

\section{Approximation of $a_{n}$}\label{AppE}

$\varphi _{n}^{(A1)}$ may be re-expressed in the following way:
\begin{eqnarray}
&\varphi _{n}^{(A1)}(z_{0})=n+1+\frac{2z_{n}^{2}}{z_{n}^{2}-1}+2\rho_{n}.z_{0}^{2} \\
&=\frac{(n+3)z_{n}^{2}-(n+1)}{z_{n}^{2}-1}+2\rho _{n}.z_{0}^{2} \\
&\varphi_{n}^{(A1)}(z_{0})=(n+1)\frac{\frac{(n+3)}{(n+1)}z_{n}^{2}-1}{z_{n}^{2}-1}+2\rho _{n}.z_{0}^{2}
\end{eqnarray}

that leads to:

\begin{eqnarray}
\begin{aligned}
&\varepsilon _{s}\varphi _{n}^{(A1)}(z_{0})-\varphi _{n}^{(A1)}(z_{s})=\\
&\varepsilon _{s}\left( (n+1)\frac{\frac{(n+3)}{(n+1)}z_{n}^{2}-1}{z_{n}^{2}-1}+2\rho _{n}.z_{0}^{2}\right) - \\
&\left( (n+1)\frac{\frac{(n+3)}{(n+1)}\varepsilon _{s}z_{n}^{2}-1}{\varepsilon _{s}z_{n}^{2}-1}+2\rho_{n}.\varepsilon _{s}.z_{0}^{2}\right)  \cr
&=\frac{(n+1)(\varepsilon _{s}-1)}{1-z_{n}^{2}}\left( 1-z_{n}^{2}\frac{\frac{(n+3)}{(n+1)}\varepsilon _{s}z_{n}^{2}-1}{\varepsilon _{s}z_{n}^{2}-1}\right)  \cr
&\varepsilon _{s}\varphi _{n}^{(A1)}(z_{0})-\varphi _{n}^{(A1)}(z_{s}) =\\
&\frac{(n+1)(\varepsilon _{s}-1)\left( f_{n}(\varepsilon_{s},z_{0})-z_{n}^{2}\right) }{(1-z_{n}^{2})f_{n}(\varepsilon _{s},z_{0})}
\end{aligned}
\end{eqnarray}

The numerator of $a_{n}$ can then be re-expressed in the following way:
\begin{eqnarray}
\begin{aligned}
&j_{n}^{(A1)}(z_{0})\left( \varepsilon _{s}\varphi _{n}^{(A1)}(z_{0})-\varphi_{n}^{(A1)}(z_{s})\right) =\cr
&\frac{(n+1)z_{0}^{n}}{(2n+1)!!}\frac{e^{-\rho
_{n}z_{0}^{2}}}{f_{n}(\varepsilon _{s},z_{0})}(\varepsilon _{s}-1)\left(
f_{n}(\varepsilon _{s},z_{0})-z_{n}^{2}\right) 
\end{aligned}
\end{eqnarray}
where the function $f_{n}$ has been defined in the article. 
The denominator can be also simplified:%
\begin{eqnarray}
\begin{aligned}
&h_{n}^{(+)}(z_{0})\left( \varepsilon _{s}\varphi _{n}^{(+)}(z_{0})-\varphi_{n}^{(A1)}(z_{s})\right) =\cr
&\frac{e^{iz_{0}}}{z_{0}^{n+1}}Q_{n}(z_{0}) \times \cr
&\left( \varepsilon _{s}\varphi _{n}^{(+)}(z_{0})-(n+1)\frac{\frac{(n+3)}{(n+1)}\varepsilon _{s}.z_{n}^{2}-1}{\varepsilon _{s}.z_{n}^{2}-1}-2\rho _{n}.\varepsilon _{s}z_{0}^{2}\right) \cr
&h_{n}^{(+)}(z_{0})\left( \varepsilon _{s}\varphi _{n}^{(+)}(z_{0})-\varphi_{n}^{(A1)}(z_{s})\right)=\cr
&\frac{e^{iz_{0}}}{z_{0}^{n+1}}\frac{Q_{n}(z_{0})}{.f_{n}(\varepsilon _{s},z_{0})}\left(\varepsilon_{s}.f_{n}(\varepsilon _{s},z_{0})g_{n}(z_{0})-(n+1)\right)
\end{aligned}
\end{eqnarray}
where $g_{n}$ has been defined in the article.
That finally leads to the following approximation of the electric Mie coefficients:
\begin{eqnarray}
\begin{aligned}
&a_{n}^{(A1)}=\frac{(n+1)z_{0}^{2n+1}}{(2n+1)!!}\frac{e^{-\rho
_{n}z_{0}^{2}-iz_{0}}}{Q_{n}(z_{0})}\times \cr
&\frac{(\varepsilon _{s}-1)\left(
f_{n}(\varepsilon _{s},z_{0})-z_{n}^{2}\right) }{\varepsilon
_{s}.f_{n}(\varepsilon _{s},z_{0})g_{n}(z_{0})-(n+1)}
\end{aligned}
\end{eqnarray}

\section{Approximation of $b_{n}$}\label{AppF}

\begin{eqnarray}
\begin{aligned}
&\varphi _{n}^{(A1)}(z_{0})-\varphi _{n}^{(A1)}(z_{s}) =\cr
&\frac{2z_{n}^{2}}{z_{n}^{2}-1}+2\rho _{n}.z_{0}^{2}-\frac{2\varepsilon _{s}z_{n}^{2}}{%
\varepsilon _{s}z_{n}^{2}-1}-2\rho _{n}.\varepsilon _{s}.z_{0}^{2} \\
&=-2\rho _{n}.z_{0}^{2}(\varepsilon _{s}-1)+2z_{n}^{2}\left( \frac{\varepsilon _{s}-1}{(z_{n}^{2}-1)(\varepsilon _{s}z_{n}^{2}-1)}\right) 
\end{aligned}
\end{eqnarray}

if we assume that $z_{n}^{2}<<1$ it then leads to: 
\begin{equation}
\varphi _{n}^{(A1)}(z_{0})-\varphi _{n}^{(A1)}(z_{s})\backsimeq (\varepsilon
_{s}-1)\left( -\frac{2z_{n}^{2}}{\varepsilon _{s}z_{n}^{2}-1}-2\rho
_{n}.z_{0}^{2}\right) 
\end{equation}

keeping the assumption $z_{n}^{2}<<1$ in the approximation $j_{n}^{(A1)}$,
it then follows that 
\begin{eqnarray}
\begin{aligned}
&j_{n}^{(A1)}(z_{0})\left( \varepsilon _{s}\varphi _{n}^{(A1)}(z_{0})-\varphi
_{n}^{(A1)}(z_{s})\right) \\
&\backsimeq \frac{z_{0}^{2n+1}}{(2n+1)!!}e^{-\rho
_{n}z_{0}^{2}}(\varepsilon _{s}-1)L_{n}(\varepsilon _{s},z_{0})
\end{aligned}
\end{eqnarray}

the denominator can be also approximated:
\begin{eqnarray}
\begin{aligned}
&\varphi _{n}^{(+)}(z_{0})-\varphi _{n}^{(A1)}(z_{s}) =\cr &\varphi
_{n}^{(+)}(z_{0})-(n+1)-\varepsilon _{s}\left( \frac{2z_{n}^{2}}{z_{n}^{2}-1}%
+2\rho _{n}.z_{0}^{2}\right)  \\
&=\varepsilon _{s}L_{n}(\varepsilon _{s},z_{0})+\varphi
_{n}^{(+)}(z_{0})-(n+1)
\end{aligned}
\end{eqnarray}

that then leads to the following approximation for $b_{n}:$%
\begin{eqnarray}
\begin{aligned}
&b_{n}^{(A1)}=\frac{z_{0}^{2n+1}}{(2n+1)!!} \times \cr &\frac{e^{-\rho
_{n}z_{0}^{2}-iz_{0}}}{Q_{n}(z_{0})}\frac{(\varepsilon _{s}-1)L_{n}(z_{0})}{%
\varepsilon _{s}L_{n}(\varepsilon _{s},z_{0})+\varphi _{n}^{(+)}(z_{0})-(n+1)%
}
\end{aligned}
\end{eqnarray}

\section{}\label{AppG}
The outgoing spherical Hankel functions can be written in the following form:
\begin{equation}\label{hn_plus}
\begin{aligned}
h^{(+)}_{n}(z) &=(-i)^{n+1}\frac{e^{iz}}{z}\sum\limits_{s=0}^{n}\frac{i^{s}}{s!(2z)^{s}}\frac{(n+s)!}{(n-s)!}\cr
&=\frac{e^{iz}}{z^{n+1}}\sum\limits_{s=0}^{n}(-1)^{n+1}\frac{i^{n+s+1}}{s!(2)^{s}}\frac{(n+s)!}{(n-s)!}z^{n-s}\cr
h^{(+)}_{n}(z) &=\frac{e^{iz}}{z^{n+1}}Q_{n}(z)
\end{aligned}
\end{equation}
where the polynomial function $Q_{n}(z)=\sum\limits_{s=0}^{n}(-1)^{n+1}\frac{i^{n+s+1}}{s!(2)^{s}}\frac{(n+s)!}{(n-s)!}z^{n-s}$

\section{}\label{AppH}
\begin{widetext}
\begin{equation}
\varepsilon_{UL}^{(e2)}=\frac{-\left(\left(\frac{n+3}{r_{n}^{2}}-2\rho_{n}^{(e)}\right)z_{0}^{2}-n\right)+\sqrt{\left(\left(\frac{n+3}{r_{n}^{2}}-2\rho_{n}^{(e)}\right)z_{0}^{2}-n\right)^{2}+4(n+1)\left(n\left(\frac{z}{r_{n}}\right)^{2}+2\rho_{n}^{(e)}\frac{z^{4}}{r_{n}^{2}}\right)}}{\left(n\left(\frac{z}{r_{n}}\right)^{2}+2\rho_{n}^{(e)}\frac{z^{4}}{r_{n}^{2}}\right)}
\end{equation}
\end{widetext}

\end{document}